# Uncertainty-Aware Transient Stability-Constrained Preventive Redispatch: A Distributional Reinforcement Learning Approach

Zhengcheng Wang, Fei Teng, *Senior Member*, *IEEE*, Yanzhen Zhou, Qinglai Guo, *Fellow*, *IEEE*, Hongbin Sun, *Fellow*, *IEEE*

*Abstract*—Transient stability-constrained preventive redispatch plays a crucial role in ensuring power system security and stability. Since redispatch strategies need to simultaneously satisfy complex transient constraints and the economic need, model-based formulation and optimization become extremely challenging. In addition, the increasing uncertainty and variability introduced by renewable sources start to drive the system stability consideration from deterministic to probabilistic, which further exaggerates the complexity. In this paper, a Graph neural network guided Distributional Deep Reinforcement Learning (GD2RL) method is proposed, for the first time, to solve the uncertainty-aware transient stability-constrained preventive redispatch problem. First, a graph neural network-based transient simulator is trained by supervised learning to efficiently generate post-contingency rotor angle curves with the steady-state and contingency as inputs, which serves as a feature extractor for operating states and a surrogate time-domain simulator during the environment interaction for reinforcement learning. Distributional deep reinforcement learning with explicit uncertainty distribution of system operational conditions is then applied to generate the redispatch strategy to balance the user-specified probabilistic stability performance and economy preferences. The full distribution of the post-redispatch transient stability index is directly provided as the output. Case studies on the modified New England 39-bus system validate the proposed method.

*Index Terms*—uncertainty, transient stability, preventive redispatch, distributional reinforcement learning, graph neural network.

## I. INTRODUCTION

With the escalating electricity demand in power systems and the increasing penetration of renewable energy sources, operational safety risks have become more prominent. Severe contingencies in power systems can cause stability deterioration, leading to major blackouts and substantial societal losses. As a crucial measure for power system stability control, preventive redispatch enables proactive adjustments of operating states in response to safety hazards, thereby enhancing the system resilience against severe contingencies [1].

The transient instability is a key problem in power system security analysis [2]. There has been significant research focus on the transient stability-constrained preventive redispatch (TSCPR) problem. This problem is commonly formulated as a transient stability-constrained optimal power flow (TSCOPF) problem, which is a complex nonlinear programming problem with constraints represented by differential algebraic equations (DAEs). In recent years, the integration of renewable energy sources and the associated uncertainty has introduced new complexities to the TSCPR, further exacerbating its challenges.

Current approaches for addressing the TSCOPF can be classified into numerical optimization methods and simplification-based methods [3]. For the former, a classic approach is to discretize the DAEs into a set of algebraic equations, which can be explicitly solved using optimization algorithms [4-6]. However, this discretization process can lead to a significant increase in the number of variables and equations [7].

Simplification-based methods usually try to approximate the complex DAE constraints with surrogate models, or simplifies the system by equivalence, whose essence lies in simplifying complex problems through approximation-based techniques. For instance, Xu *et al*. employ the one-machine-infinite-bus equivalence technique on multi-machine systems and utilize trajectory sensitivity analysis to convert the intricate DAE constraints into linear ones. Subsequently, they solve the simplified TSCOPF using linear programming [8]. To enhance the accuracy of approximating transient stability constraints, several data-driven surrogate models have been proposed [9-13]. Zhou *et al*. use a two-stage support vector machine to approximate the transient stability constraints and the chaotic particle swarm optimization is applied to solve the TSCOPF. Zhang *et al*. introduce an XGBoost-based transient stability classifier as the surrogate model and employ the differential evolution algorithm for optimization [10]. Su *et al*. utilize a deep belief network as the surrogate model and leverage the reference-point-based nondominated sorting genetic algorithm for optimization [11]. In reference [12], the Bayesian neural network is employed for stability prediction, and the TSCOPF is solved by Bayesian optimization. Fu *et al*. adopt the non-parametric kernel regression for critical clearing time prediction, followed by model reduction and linearization to transform the preventive control problem into a mixed-integer linear program [13]. Although these methods can improve the computational speed, they often suffer from a lack of accuracy and the online optimization part still takes significant time.

Deep reinforcement learning (DRL), as an emerging approach for solving complex decision-making problems, has been applied to TSCPR problem [7, 14]. Zeng *et al*. employ the distributed deep deterministic policy gradient (D3PG) algorithm to tackle transient

This work was supported by National Natural Science Foundation of China (Authorized Number: U22B2097). (*Corresponding author: Fei Teng and Hongbin Sun.*)

Z. Wang, Y. Zhou, Q. Guo and H. Sun are with the State Key Laboratory of Power Systems, Department of Electrical Engineering, Tsinghua University, Beijing 100084, China (e-mail: wzc20@mails.tsinghua.edu.cn; zhouyzh@126.com; guoqinglai@tsinghua.edu.cn; shb@tsinghua.edu.cn).

F. Teng is with the Department of Electrical and Electronic Engineering, Imperial College London, SW7 2AZ London, U.K (e-mail: f.teng@imperial.ac.uk).

stability preventive control, where the time-domain simulation (TS) is integrated into the environment to compute the reward at each step [7]. However, due to the inclusion of TS in the environment for transient stability assessment (TSA), the training process can be time-consuming. In reference [14], a graph convolutional neural network is trained for TSA and transferred to DRL to enhance learning efficiency. DRL-based methods can provide effective redispatch strategies in several seconds and is a promising method for TSCPR even in some emergency cases.

In recent years, the rise of renewable energy has introduced variability and uncertainties to the stability analysis. Consequently, the stability index will transfer from deterministic to probabilistic. Since online preventive redispatch focus on the forward-looking time scope, it is essential to take into consideration the variability and uncertainty associated with the stability criterion during the time interval. However, for the previous approaches, the numerical optimization methods and simplification-based methods cannot take uncertainty into account. Although the online reinforcement learning (RL) has the ability to adapt to such uncertainty through continuous online exploration and correction, its efficiency is extremely low. Moreover, since RL operates on expected values, it cannot directly process a distribution.

When addressing the TSCOPF with uncertainty, the objectives and constraints need to be extended into an uncertain domain, which enlarges the solution space and poses challenges for analytical analysis. As a result, solving the uncertainty-aware TSCOPF becomes even more difficult. References [15]-[18] try to address TSCOPF with uncertainty. References [15] and [16] make simplifications by trajectory sensitivity analysis and equivalence techniques, however, due to the use of trajectory sensitivity and equivalence, these two works lost some accuracy. Besides, they focused on robust target, and unable to consider the complete post-control distribution. Reference [17] only produces raw moments from their point estimate method. Since the Gaussian process is utilized, reference [18] needs to make normality assumptions on the probability distributions of the output variables. These approaches lead to loss of information and biases.

In this work, a graph neural network-guided distributional deep reinforcement learning (GD2RL) framework is proposed to address the online uncertainty-aware TSCPR (U-TSCPR) problem for future scenarios. A heterogeneous message passing graph neural network (HMPNN) is trained as a surrogate transient simulator to efficiently and accurately characterize transient stability constraints; A novel distributional deep reinforcement learning (D2RL) framework is designed to optimize one single redispatch strategy for an uncertain operating state, while simultaneously obtaining the distribution of post-redispatch transient stability index (TSI); The HMPNN-based transient simulator is integrated into the D2RL environment to enhance the training efficiency.

The remainder of this paper is organized as follows. In Section II, the statement of the uncertainty-aware TSCPR problem is described and then the framework of GD2RL is introduced. In Section III, we delve into the construction of the HMPNN-based transient stability simulator. The detailed design of GD2RL framework for solving the U-TSCPR will be introduced in Section IV. The model structure of GD2RL agents as well as the training process are discussed in Section V. To validate the effectiveness of the proposed method, comprehensive case studies are conducted on the modified New England 39-bus system, which is discussed in Section VI. Finally, Section VII summarizes our findings and draw conclusions.

## II. PROBLEM STATEMENT & THE METHODOLOGICAL FRAMEWORK

In this section, the mathematical formulation of the U-TSCPR problem is described. Additionally, the overview of the GD2RL framework is introduced, highlighting its key components and the overall methodology employed in solving the U-TSCPR problem.

### A. Mathematical formulation of the uncertainty-aware transient stability-constrained preventive redispatch

In our settings, the uncertainty of renewable energy sources is considered. On this premise, the U-TSCPR problem can be mathematically described as an uncertainty-aware TSCOPF (U-TSCOPF) problem as shown in (1)-(7):

$$\min_{\Delta P_g, g \in \Gamma_a} \left( \lambda_c \sum_{g \in \Gamma_a} c_g \Delta P_g + \lambda_s p(TSI_\Delta < 0) \right) \quad (1)$$

s.t.

$$P_g = P_g^0 + \Delta P_g, \forall g \in \Gamma_a \quad (2)$$

$$p(\underline{V_i} < V_i < \overline{V_i}) > \gamma_V, \forall i \in \mathcal{N} \quad (3)$$

$$p(\underline{P_g} < P_g < \overline{P_g}) > \gamma_P, \forall g \in \Gamma_a \quad (4)$$

$$P_r \sim \varphi_r, \forall r \in \Gamma_r \quad (5)$$

$$G(\boldsymbol{c}, \boldsymbol{s}, \boldsymbol{z}) = 0 \quad (6)$$

$$TSI_\tau = (360° - |\Delta \delta_\tau|_{max}) / (360° + |\Delta \delta_\tau|_{max}), \tau \in \mathcal{T} \quad (7)$$

The objective function (1) aims to achieve a balance between stability and economy. It consists of two terms: the redispatch cost term and the transient stability risk term. In (1), the redispatch cost term is determined by the adjustment cost for each controllable generator. $\Gamma_a$ denotes the set of controllable generators, typically referring to conventional generators excluding renewable sources. $\Delta P_g$ represents the amount of active power redispatch for generator $g$, and $c_g$ denotes the corresponding adjustment cost. The weight for the redispatch cost term is denoted as $\lambda_c$.

The second term in the objective function quantifies the risk of transient instability after redispatch. $TSI_\Delta$ represents the TSI after applying the redispatch strategy $\Delta$. The term $p(TSI_\Delta < 0)$ represents the probability of the uncertainty state becoming transient unstable after redispatch. The weight for the transient stability risk term is denoted as $\lambda_s$.

Inequalities (3) and (4) represent the probability constraint for bus voltage violation and generator active power output violation respectively, in which $\gamma_V$ and $\gamma_P$ represent the probability thresholds, and $\mathcal{N}$ denotes the set of buses. Constraint (5) represents the power of renewable source $P_r$ obeys a certain distribution $\varphi_r$, which can be statistically obtained from historical measurement and historical forecast data. The set of renewable units is denoted as $\Gamma_r$. Equation (6) represents a set of equality constraints including the power flow constraints and the power balance constraints. The control variables are denoted by $\boldsymbol{c}$, the state variables by $\boldsymbol{s}$, and the topological variables by $\boldsymbol{z}$. The $TSI_\tau$ is defined in (7) according to [19], in which $|\Delta \delta_\tau|_{max}$ means the maximum post-contingency rotor angle difference among generators after contingency $\tau$ occurs, and $\mathcal{T}$ denotes the contingency set. TSI>0 indicates the system is stable after contingency and vice versa.

Solving the U-TSCOPF problem stated in equations (1)-(7) directly poses significant challenges, as it involves solving DAEs implied in constraint (7). Moreover, the presence of uncertainty further complicates the analytical solution. To address these difficulties, the framework of GD2RL is proposed as a solution to the U-TSCOPF problem.

*B. The framework of the GD2RL*

The flowchart illustrating the GD2RL framework is depicted in Fig. 1, and it consists of three stages:

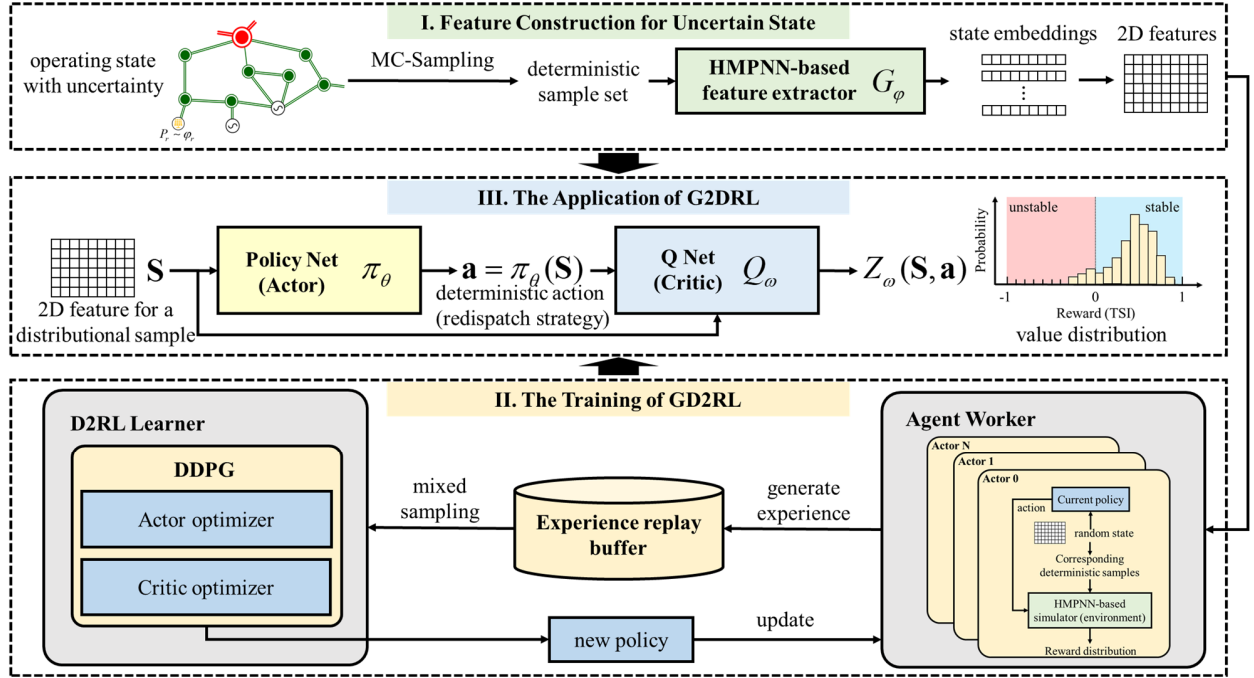

Fig. 1 The flowchart of GD2RL.

*Stage I: Feature Construction for Uncertain State.* First, the Monte-Calo (MC) sampling is performed on an uncertain state to generate a series of deterministic operating states. Subsequently, an HMPNN-based feature extractor is trained to convert these deterministic operating states into embedding vectors. These vectors are then concatenated into a 2D feature array, serving as a representation of the initial uncertain state.

*Stage II: The Training of GD2RL.* During the training of GD2RL, the concept of a distributional sample and a deterministic sample are defined. A distributional sample refers to an operating scenario with uncertainty, while a deterministic sample corresponds to a deterministic power flow state sampled from the distributional sample. The training of GD2RL can be divided into exploring and learning. The exploring is performed by the agent worker, where the HMPNN-based simulator is utilized to serve as the environment. The base state of a distributional sample is randomly sampled from the power system, and the distributional information of the power flow variables are obtained from historical data. The feature construction is performed as in Stage I, and the resulting 2D feature array is input into the current policy network to generate the deterministic action. Simultaneously, the corresponding deterministic samples sampled from the uncertain state are input into the HMPNN-based simulator to obtain the reward distribution. The experiences containing random states, actions and reward distributions are then added to the experience buffer. To expedite exploration, a distributed parallel technique is applied within the agent worker. During learning, the distributional deep deterministic policy gradient (D3PG) [20] is adopted to train the policy. Experiences are sampled from the replay buffer, and the D2RL learner update the actor and critic through DDPG. The newly updated policy replaces the old policy in the agent worker.

*Stage III: The Application of GD2RL.* Once the training process is complete, the policy net and the critic net (also known as Q net) can be used for online application. The random operating state is input into the policy net to obtain the deterministic strategy. Additionally, the random state and action are fed into the critic, outputting the value distribution. This value distribution directly reflects the TSI distribution, providing comprehensive information to assess the operating state and redispatch strategy.

III. THE HMPNN-BASED TRANSIENT STABILITY SIMULATOR

In the framework presented in Section II, the characterization of operating states that involve uncertainty is achieved through the utilization of MC sampling. However, evaluating the impact of actions on these sampled points using traditional TS within the RL environment can be really time-consuming. To address this challenge, an HMPNN-based transient stability simulator is proposed to perform the pseudo transient simulation.

As a graph neural network, MPNN provides a means to propagate node and edge information across the graph [21]. MPNN is particularly well-suited for modeling the power systems, as it captures the similarities between information propagation in the model and energy propagation in the power system. This often leads to improved performance compared to traditional data-driven methods in TSA tasks. Furthermore, MPNN exhibits better generalization capabilities when topological change occurs [22]. In this paper, the power system is abstracted into a graph where the nodes are divided into generator nodes (gen-nodes) and other nodes (other-nodes), so the MPNN is expanded to an HMPNN with these heterogeneous nodes. The overall framework of the HMPNN-based simulator is shown in Fig. 2.

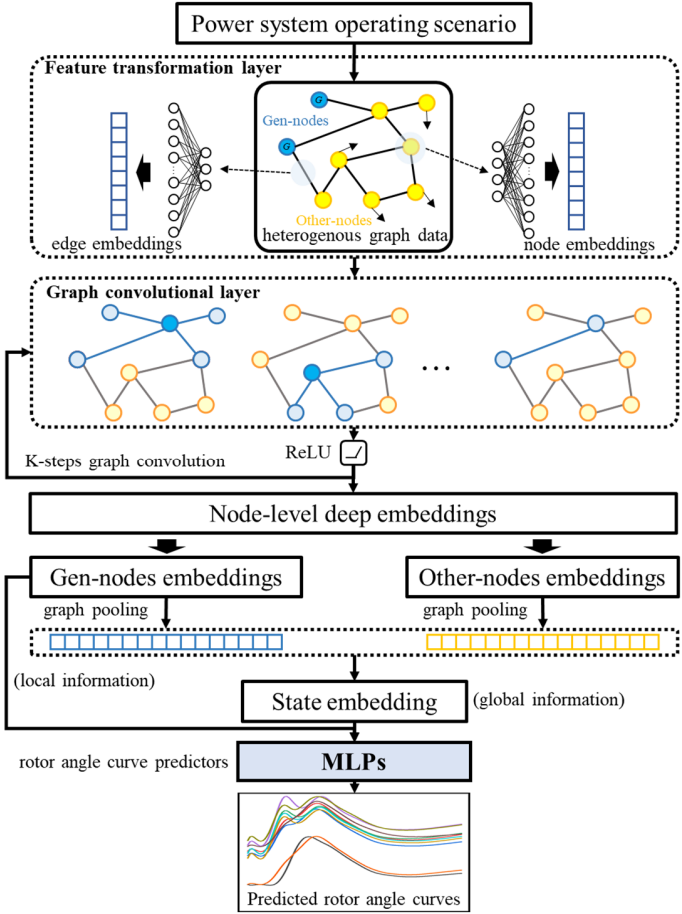

Fig. 2 The framework of HMPNN-based transient simulator

In this framework, a deterministic sample is abstracted as a heterogeneous graph. For the gen-nodes, the original node features can be represented as:

$$v_{gen} = [P_g, P_L, Q_g, Q_L, V, \theta] \quad (8)$$

which contains the active and reactive power on generator and load of this bus, as well as the magnitude and phase angle of the bus voltage. For the other-nodes, features are as follows:

$$v_{other} = [P_L, Q_L, V, \theta] \quad (9)$$

For each edge in the graph, the feature vector is:

$$v_{egde} = [P_e, Q_e, F] \quad (10)$$

in which $F$ is a 0-1 variable representing whether there is an expected fault. Given the presence of two distinct types of nodes and the utilization of a bidirectional structure, the edges can be classified into five categories, namely: "gen2other", "other2gen", "other2other", "reverse_gen2other", and "reverse_other2gen".

In the feature transformation layer, the original features on each node and edge are transformed into embeddings via single hidden layer neural networks. Notably, the network parameters are shared separately for each type of node and edge. Subsequently, the heterogeneous graph with node and edge embeddings is inputted to a graph convolutional (GC) layer, where a K-steps GC is performed to pass and aggregate information on the graph. In this paper, K is set to be two.

Following the GC, the information on the graph is presented as node-level deep embeddings, which can be divided into gen-nodes embeddings and other-nodes embeddings. Node embeddings with graph structure are then transformed into vector-form embeddings via a graph pooling layer, and the vector-form embeddings from gen-nodes and other-nodes are concatenated to get the state embedding. This state embedding encapsulates the global information of the original operating scenario. Thus, the proposed model can function as an HMPNN-based feature extractor, with the state embedding serving as its middle output.

Instead of only predicting the transient stability or the TSI, the proposed simulator takes the post-contingency rotor angle curve for a generator as the predicting target, and the real rotor angle curves of generators obtained from the TS is used as the labels during the training stage. A multi-layer perceptron (MLP) [23] is constructed for each gen-node, which takes both the state embedding and the gen-node embedding associated with the specific node as input. The inclusion of the gen-node embedding in the input is motivated by its ability to capture the local information surrounding the node. By incorporating the gen-node embedding, the features specific to the local area surrounding the generator can be emphasized. The combination of global and local information proves beneficial for the angle curve prediction task.

During the data pre-process stage, all features except fault information $F$ are normalized by Z-score normalization:

$$\hat{X}_i = (X_i - \bar{X}_i)/\sigma_{X_i} \quad (11)$$

where $\bar{X}_i$ is the mean for feature $X_i$ and $\sigma_{X_i}$ is the standard deviation. However, this normalization method is not suitable for processing the angle curve labels due to the significant differences between stable and unstable curves, and the potentially large values of rotor angles in unstable scenarios. To address this issue, a logarithm-based data pre-processing method is proposed, which can be expressed as (12):

$$\hat{y}_i = \begin{cases} \ln(y_i), & y_i > 1, \\ -\ln(-y_i), & y_i < -1, \\ 1, & -1 \le y_i \le 1 \end{cases} \quad (12)$$

And the corresponding reverse transformation is:

$$y_i = \begin{cases} e^{\hat{y}_i}, & \hat{y}_i \ge 0, \\ -e^{-\hat{y}_i}, & \hat{y}_i < 0 \end{cases} \quad (13)$$

When $y_i$ ranges between -1 and 1, (12) is a lossy transformation. However, the absolute magnitude of rotor angles tends to be large, and the specific variations within a few degrees are not of significant concern.

The HMPNN-based simulator is trained by supervised learning. The output dimension of MLPs is 100, representing the post-contingency dynamic process within ten seconds. By leveraging the HMPNN-based simulator, the prediction of post-contingency angle curves for generators can be achieved rapidly given the steady-state input and fault information. Unlike methods that solely predict stability or TSI, the proposed simulator provides a comprehensive depiction of the dynamic process. Consequently, it offers improved interpretability and allows the operators to assess the reliability of the predicted result based on their operating experience. Furthermore, the predicted angle curves facilitate operators in analyzing the post-contingency characteristics of the operating scenario and evaluating the effectiveness of subsequent redispatch strategies.

IV. THE GD2RL FRAMEWORK FOR SOLVING U-TSCPR

In this section, the procedure of utilizing the GD2RL framework to address the U-TSCPR problem is presented. First, the concept of distributional reinforcement learning (DistRL) is introduced. Then, basic notations of formulating the U-TSCPR problem into the Markov decision process is addressed.

*A. Distributional Reinforcement Learning*

In the setting of traditional RL, an agent should aim to maximize its expected value $Q$ [24]. The traditional Bellman operator $\mathcal{B}_\pi$ is used to describe such value in terms of the expected reward and expected next-state outcome [25]:

$$\mathcal{B}_\pi Q^\pi(s,a) := \mathbb{E}[R(s,a)] + \gamma \mathbb{E}_{P,\pi}[Q^\pi(s',a')] \quad (14)$$

with state $s$, action $a$, policy $\pi$, reward $R$ and discount $\gamma$. $Q$ means the value function, and $P$ is the state transition function. However, when employing traditional RL, a significant portion of the distribution information is lost, leaving only the expectation as the remaining information. To resolve this shortcoming, the DistRL is proposed in [26] that explicitly captures the entire value distribution instead of the expectation alone. This framework allows for a more comprehensive representation of uncertainties and has demonstrated promising results. In their work, the C51 algorithm is proposed to transform the deep Q-learning into a distributional form. Reference [20] and [27] combine the DistRL with the deterministic policy gradient algorithm.

The core of DistRL is the distributional Bellman operator $\mathcal{T}_\pi$:

$$\mathcal{T}_\pi Z^\pi(s,a) := R(s,a) + \gamma Z^\pi(s', \pi(s') | s, a) \quad (15)$$

where $Z$ is the value distribution.

Our work leverages the DistRL framework to tackle the uncertainty inherent in the operating state itself. By adopting DistRL, we aim to capture and model the distribution of potential outcomes in the uncertain operating state, enabling more effective decision-making under uncertainty. Since the uncertainty of the state itself is considered, formula (15) is then transformed into:

$$\mathcal{T}_\pi Z^\pi(\mathbf{S},\mathbf{a}) := R(\mathbf{S},\mathbf{a}) + \gamma Z^\pi(\mathbf{S}', \pi(\mathbf{S}') | \mathbf{S}, \mathbf{a}) \quad (16)$$

where $\mathbf{S}$ is the state distribution and $\mathbf{a}$ means the action vector.

Actually, traditional online RL can also enable the agent to progressively uncover the patterns of uncertainty in renewable sources through continuous exploration. However, in practical operations, there is usually some prior knowledge available regarding the distribution of renewable sources. By directly embedding this prior knowledge in DistRL, the exploration space for the agent can be substantially diminished, resulting in enhanced algorithmic learning efficiency.

*B. Designing of the GD2RL framework for U-TSCPR*

To tackle the U-TSCPR problem via DistRL, basic notations are defined to form the Markov decision process of the preventive redispatch. Since the future state of the power systems can be regarded as only related to the current state and the action due to the short time scale for deriving redispatch strategies, the preventive redispatch problem can rigorously satisfy the Markov property. An actor-critic structure [28] is adapted to perform the GD2RL, where the actor (policy) network aims to find the optimal action under certain state and the critic ($Z$) network learns to estimate the reward for the given action and state. The key components of the GD2RL for U-TSCPR is shown in Fig. 3.

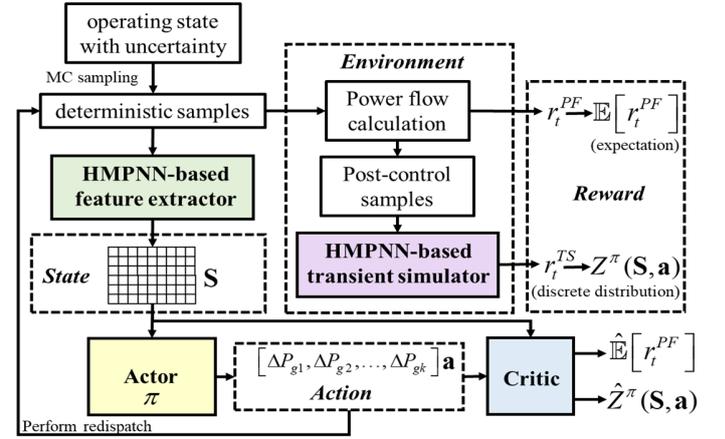

Fig. 3 The key components of the GD2RL for U-TSCPR.

*State*. In traditional RL, states are typically composed of power flow variables. However, in our problem, the state is replaced by state distribution. Therefore, deterministic states are first sampled from the distributional sample and then the HMPNN-based feature extractor is applied to extract the state embeddings sample point. The concatenated matrix of state embeddings is then considered as the state distribution.

*Action*. The objective of agents is to optimize the active power outputs of the generators. However, there are certain limitations on the generator outputs. Specifically, the generator connected to the slack bus is responsible for balancing the power imbalance, and thus its active power output cannot be adjusted. Additionally, the outputs of renewable sources are also hard to modify unless measures like resource curtailment are utilized, which are generally discouraged. In our setting, the action space is defined as the active power adjustments of the adjustable generators, and the magnitude of these adjustments in one step is constrained within [$-a$, $a$] MW, where $a$ is set to 50 in this paper.

*Reward*. The reward is the feedback received by the agent, based on which the agent learns to generate redispatch strategies. First, the value of a single "deterministic state-action pair" is designed, which is divided into two parts:

$$v_t^{PF} = \begin{cases} \psi_{\min}, & \text{if power flow diverged} \\ \lfloor G_{PF}(s_t) \rfloor_{\psi_{\min}}, & \text{else} \end{cases} \quad (17)$$

$$v_t^{TS} = TSI(s_t) \quad (18)$$

$v_t^{PF}$ represents the value for the post-redispatch power flow state, where $G_{PF}$ is the penalty function for over-limit of active power and nodal voltage, which is shown as:

$$G_{PF}(s_t) = R_P(s_t) + R_V(s_t) \quad (19)$$

$$R_P(s_t) = -\sum_g (\underline{P}_g - P_g) \cdot \mathbf{1}_{P_g < \underline{P}_g} \\ -\sum_g (P_g - \overline{P}_g) \cdot \mathbf{1}_{P_g > \overline{P}_g} \quad (20)$$

$$R_V(s_t) = -\sum_i (\underline{V}_i - V_i) \cdot \mathbf{1}_{V_i < \underline{V}_i} \\ -\sum_i (V_i - \overline{V}_i) \cdot \mathbf{1}_{V_i > \overline{V}_i} \quad (21)$$

$$\mathbf{1}_{condition} = \begin{cases} 1, & \text{if condition is True} \\ 0, & \text{if condition is False} \end{cases} \quad (22)$$

$v_t^{TS}$ represents the post-redispatch TSI which ranges within [-1, 1]. In (17), $\Psi_{min}$ denotes the lower limit of the penalty function, which is set to -1. Based on the value for each deterministic sample, the reward distribution for each step can be defined as:

$$r_{t+1}^{PF} = \mathbb{E}\left(v_{t+1}^{PF}\right) - \mathbb{E}\left(v_t^{PF}\right) \quad (23)$$

$$Z_{t+1}^{TS} = I^{v_{t+1}^{TS}} - I^{v_t^{TS}} = \left[p_{z_0}^{t+1} - p_{z_0}^t, p_{z_1}^{t+1} - p_{z_1}^t, ..., p_{z_A}^{t+1} - p_{z_A}^t\right] \quad (24)$$

where $I$ denote the TSI distribution which is represented by a discrete distribution based on the MC samples, with the number of bins $A$ set to be 51 [26]. $z_i$ is the support of the interval which can be represented by a set of atoms $\{z_i = -1 + i\Delta z: 0 \leq i < 51\}$, and $p$ denotes the probability of the corresponding support. Therefore, the critic net in our framework has an output dimension of 52. The first output corresponds to the expectation term of $r_t^{PF}$, while the remaining represents the probabilities of the random variables $z_i$. Rather than being included in the reward, the redispatch cost is considered in the loss function as a separate item, which will be introduced in the next part.

*Environment*. The environment is composed of the power flow calculation and the HMPNN-based simulator. As illustrated in Fig. 3, once the action is generated by the actor, it is applied to the deterministic samples obtained through MC sampling. First, the redispatch strategy is applied to the deterministic samples, leading to a new power flow state. This is achieved by employing the Newton method, which efficiently calculates the updated power flow considering the effects of the action. Subsequently, the post-redispatch angle curves of each deterministic samples are predicted using the HMPNN-based simulator. Based on the result of power flow calculation and the pseudo transient simulation on each deterministic samples, $r_t^{PF}$ and $r_t^{TS}$ for the distributional sample can be derived by aggregating them together. Due to the high prediction accuracy of the HMPNN-based simulator, the post-redispatch TSI distribution obtained via the pseudo transient simulation is also close to the real one. Since both the power flow calculation and the pseudo transient simulation can be completed within a short period of time, the time consumed by the GD2RL algorithm in the environment interaction stage will be significantly less compared to traditional RL.

During the application phase, the deterministic operating states will be sampled firstly from the target uncertain scenario that requires redispatch. Then, the HMPNN-based feature extractor is used to extract the state embeddings of these deterministic operating states, which are concatenated as a matrix to represent the state distribution of the target scenario. The state distribution matrix will be input into the trained actor network of the GD2RL to obtain the redispatch action. To limit the adjustment magnitude of the GD2RL agent, a step-wise approach is adopted during the redispatch process. After the actor outputs an action, the target scenario will be adjusted according to the action in the simulation environment, and the TSI distribution outputted by the critic can be directly observed. If the distribution meets the predefined safety criterion, the redispatch process ends; otherwise, the adjusted scenario is constructed as a new state distribution, input into the actor, and a new action is obtained. When the specified maximum number of steps is reached and the TSI distribution still does not meet the requirements, a failure signal will be outputted to warn the operators for manual intervention.

## IV. THE MODEL STRUCTURE AND THE TRAINING PROCESS OF GD2RL AGENTS

The previous section introduced the GD2RL framework used for the U-TSCPR problem. In this section, the specific model structure of agents in the GD2RL, which means the detailed implementation of the actor and critic, will be described, and then the design of the loss function and the corresponding training algorithm are introduced.

### A. Model structure of the actor and the critic

Both the actor and the critic are composed by two parts: a convolutional network and a task network. Since the state distribution is formulated as a 2D array whose size varies with the number of deterministic samples, the convolutional network is chosen to extract deep features, while the task network is responsible for action prediction (for actor) and reward evaluation (for critic). The structure of actor and critic is shown in Fig. 4.

The parameters of the convolutional networks for the actor and the critic are shared. This allows consistency and coherence between two networks. For the task networks, MLP is adopted, and the actor and critic networks employ identical hidden layer structures within their respective task networks according to [28]. The output of actor is a vector of length equal to the size of the action space, which is nine, and the output of the critic is 52 as has been introduced in the reward designing part.

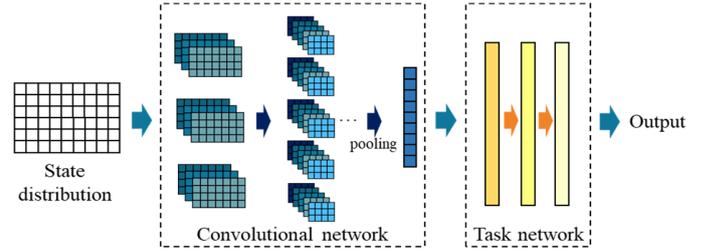

Fig. 4 The model structure of the actor and the critic.

### B. Training of the GD2RL agent

For the proposed GD2RL agent, it tries to maximize the discounted return in an episode as follows:

$$\max G = \mathbb{E}\left(Z_{post}^{TSI}\right) + Z_{post}^{PF} - \mathbf{c} \cdot \sum_{t=0}^{T} \mathbf{a}_t^{\top} \quad (25)$$

$$Z_{post}^{TSI} = I^{v_0^{TS}} + \sum_{t=0}^{T} Z_{t+1}^{TS} \quad (26)$$

$$Z_{post}^{PF} = \sum_{t=0}^{T} r_{t+1}^{PF} \quad (27)$$

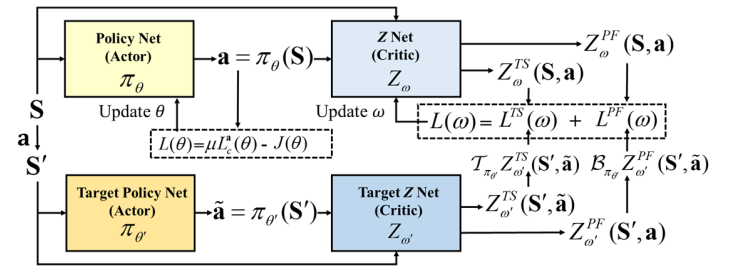

Fig. 5 The training framework of GD2RL based on D3PG.

Formula (26) represents the post-redispatch TSI distribution after the episode, where the first term means the pre-redispatch TSI distribution. Formula (27) means the post-redispatch penalty for the power flow violation, and the third term in (25) is the redispatch cost, where $\mathbf{c}$ is the cost vector for adjustable generators and $\mathbf{a}_t$ is the action for step $t$. In this paper, the D3PG algorithm is adopted to train the agent. The D3PG trains the actor and critic networks and their corresponding target networks, as shown in Fig. 5.

In the D3PG algorithm, the loss function employed for updating the weights $\omega$ of the critic is composed of two parts:

$$L(\omega) = L^{TS}(\omega) + L^{PF}(\omega) \quad (28)$$

$$L^{TS}(\omega) = \mathbb{E}_\rho \left[ d\left( Z^{TS}_\omega(\mathbf{S},\mathbf{a}), \mathcal{T}_{\pi_{\theta'}} Z^{TS}_{\omega'}(\mathbf{S}',\tilde{\mathbf{a}}) \right) \right]$$
$$= \mathbb{E}_\rho \left[ D_{KL}\left( Z^{TS}_\omega(\mathbf{S},\mathbf{a}) \| \mathcal{T}_{\pi_{\theta'}} Z^{TS}_{\omega'}(\mathbf{S}',\pi_{\theta'}(\mathbf{S}')) \right) \right] \quad (29)$$

$$L^{PF}(\omega) = \mathbb{E}_\rho \left[ \left\| Z^{PF}_\omega(\mathbf{S},\mathbf{a}) - \mathcal{B}_{\pi_{\theta'}} Z^{PF}_{\omega'}(\mathbf{S}',\pi_{\theta'}(\mathbf{S}')) \right\|^2 \right] \quad (30)$$

where $L^{TS}(\omega)$ represents the loss from $Z^{TS}_t$, and $D_{KL}$ means the KL divergence. $L^{PF}(\omega)$ denotes the loss from $r^{PF}_t$. $\rho$ means samples sampled from the experience replay buffer. The loss function to update the weights $\theta$ of the actor also has two parts:

$$L(\theta) = \mu L^{\mathbf{a}}_c(\theta) - J(\theta) \quad (31)$$

$$L^{\mathbf{a}}_c(\theta) = \mathbb{E}_\rho \left[ \pi_\theta(\mathbf{S}) \cdot \mathbf{c}^\top \right] \quad (32)$$

$$J(\theta) = \mathbb{E}_\rho \left[ \mathbb{E}(Z^{TS}_\omega(\mathbf{S},\pi_\theta(\mathbf{S}))) + Z^{PF}_\omega(\mathbf{S},\pi_\theta(\mathbf{S})) \right] \quad (33)$$

where $L^{\mathbf{a}}_c(\theta)$ is the loss term for the redispatch cost and $\mu$ denotes the weight for the redispatch cost term, which is set to 0.1. $J(\theta)$ means that the actor is updated to maximize the expected return of the critic output. The training targets $L(\omega)$ and $L(\theta)$ imply that the GD2RL aims to learn a policy that can maximize the expected reward and minimize redispatch cost, while making the predicted TSI reward distribution and the actual post-redispatch TSI distribution to be as close as possible.

To accelerate the exploration process in the DistRL, a distributed parallel technique for the environment is applied, which changes the D3PG algorithm into D4PG. The detailed training process is shown in Algorithm 1 in Appendix A.

In the training stage, a mixed sampling strategy is adopted when sampling mini-batch from the replay buffer. During the power system operation, some scenarios are hard to redispatch, while others are easier. The goal of the policy network is to achieve good redispatch performance for both challenging and easy scenarios. Therefore, random sampling and importance sampling are performed alternately, resulting in the mixed sampling strategy. Random sampling entails equal probabilities for sampling all distributional scenarios, ensuring a fair representation of the entire distribution. The importance sampling assigns higher probabilities to scenarios with lower rewards. This approach enables the model to pay more attention to challenging scenarios. By alternating between random sampling and importance sampling, the model achieves a balanced exploration of the sample space, enhancing its adaptability to a wide range of scenarios.

## VI. CASE STUDY

To validate the effectiveness of the proposed method, the modified New England 39-bus system is used to perform the numerical experiments. All case studies are performed on a computing platform with Intel(R) Xeon(R) E5-2640v4 @ 2.40 GHz, 128 GB RAM and NVIDIA Tesla V100S. All models are implemented using PyTorch and PyTorch-Geometric.

### A. The dataset generation for the transient simulator training

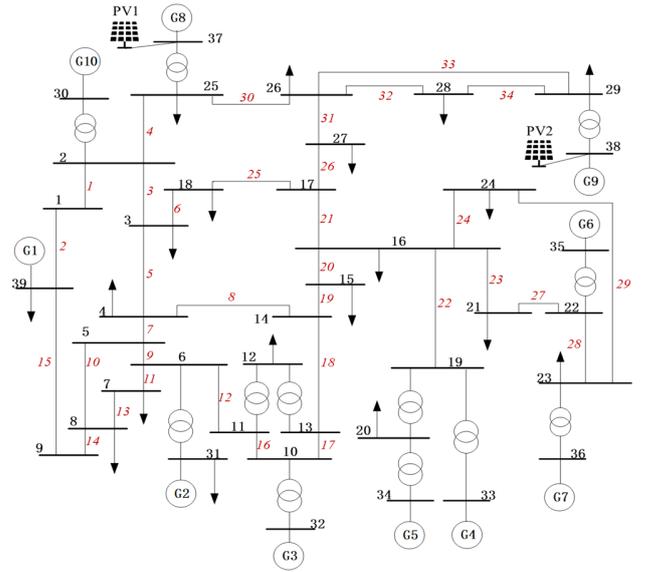

Fig. 6 The modified New England 39-bus system.

In the testing system shown in Fig. 6, two photovoltaic units are added to the system, and G2 serves as the balance machine. PV1 is connected to bus 37 and PV2 is connected to bus 38. The redispatch cost for each generator in the modified New England 39-bus system is determined according to [29], as Table I shows:

TABLE I REDISPATCH COST FOR EACH GENERATORS

| Gen. | G1 | G2(B) | G3 | G4 | G5 |
|---|---|---|---|---|---|
| Redispatch Cost $/MW | 6.0 | / | 4.5 | 9.0 | 9.0 |
| Gen. | G6 | G7 | G8 | G9 | G10 |
| Redispatch Cost $/MW | 6.0 | 1.0 | 5.0 | 6.0 | 2.0 |

During the dataset generation, the overall generation and load are uniformly divided into 9 levels, ranging from 80% to 120%. For each generation and load level, random fluctuations within ±10% are added to each generator and load. Besides, a disturbance within ±5% is applied to the voltage of each generator. For each AC line in the system except for AC line 22 (which will split the system once being removed), a three-phase permanent short-circuit fault at 50% of the line is set in turn, and the fault clearing time is 0.1s. For each generation and load level, 500 samples are generated, and a total number of 33×9×500=148500 samples are generated, in which about 37% samples are unstable. The post-contingency angle curves within ten seconds for G1-10 are recorded as labels.

The dataset is divided into the training, validation and testing set according to the "5:3:2" principle. During the training, the mean square error is used as the loss function and Adam [30] is chosen as the optimizer. The optimal hyper-parameters of the HMPNN-based simulator are selected based on the validation performance. The dimension of the node and edge embeddings in the GC layers is set to 30. For the MLPs, a two hidden layer structure is adopted, with the number of hidden neurons being 200 and 150 respectively. The initial learning rate is set to 0.01 and the learning rate is decayed by 90% every 10 epochs. The batch size is set to 512, and an early-stop technique is employed.

## B. Testing for the HMPNN-based transient simulator

To validate the effectiveness of the HMPNN-based transient simulator, the mean prediction error for the curves (MPEC) is used as an evaluating index. Since the stability of a scenario can also be obtained from the predicted curves, F1 score (F1), accuracy (Acc.), true negative rate (TNR) and true positive rate (TPR) are also used to evaluate the performance of the simulator. The index definitions are listed as follows:

$$MPEC = \frac{1}{N_T N_c} \times \sum_{c=1}^{C} \sum_{t=1}^{T} \frac{|y_{c,t} - \hat{y}_{c,t}|}{|y_{c,t}|} \times 100\% \quad (34)$$

$$TNR = 100\% \times TN/(TN+FP) \quad (35)$$

$$TPR = 100\% \times TP/(TP+FN) \quad (36)$$

$$Acc. = 100\% \times (TN+TP)/(TN+TP+FN+FP) \quad (37)$$

$$F1 = 100\% \times (2 \times TP)/(2 \times TP+FP+FN) \quad (38)$$

where $N_T$=100 denotes the length of a curve, $N_C$=10 means the number of curves, $y_{c,t}$ represents the rotor angle label of generator $c$ at timestep $t$ after reverse transformation via (13). TP stands for true positive, representing correctly classified unstable samples. TN stands for true negative, representing correctly classified stable samples. FP stands for false positive, representing falsely classified unstable samples. FN stands for false negative, representing falsely classified stable samples.

The model performance on the testing set is shown in Table II, and the MLP-based simulator is used for comparison. From the result it can be seen that the HMPNN-based simulator achieves an MPEC of 5.67%, which is far lower than the MLP-based one, and it also has better transient stability classification performance.

TABLE II COMPARISON BETWEEN MLP AND HMPNN-BASED SIMULATOR

| Index | MLP-based | HMPNN-based |
|---|---|---|
| MPEC | 12.67% | **5.67%** |
| F1 | 98.64% | **99.15%** |
| Acc. | 99.00% | **99.37%** |
| TNR | 99.20% | **99.48%** |
| TPR | 98.66% | **99.18%** |

Some examples of the predicted angle curves for generators from the HMPNN-based simulator are shown in Fig. 7. Figures on the left are simulation results from TS while the ones on the right display results from the HMPNN-based simulator.

The presented figures suggest that the HMPNN-based transient simulator can accurately capture the post-contingency dynamic characteristics of generator rotor angles, irrespective of whether the system is in a stable or unstable condition. Additionally, the total time consumption of conducting simulation using TS and HMPNN-based simulator on 1000 samples are tested. The TS took 436 seconds in total, while the latter only took 0.39 seconds. This indicates that the HMPNN-based simulator can well approximate the traditional time-consuming TS in addressing the preventive redispatch problem, with far better computational efficiency.

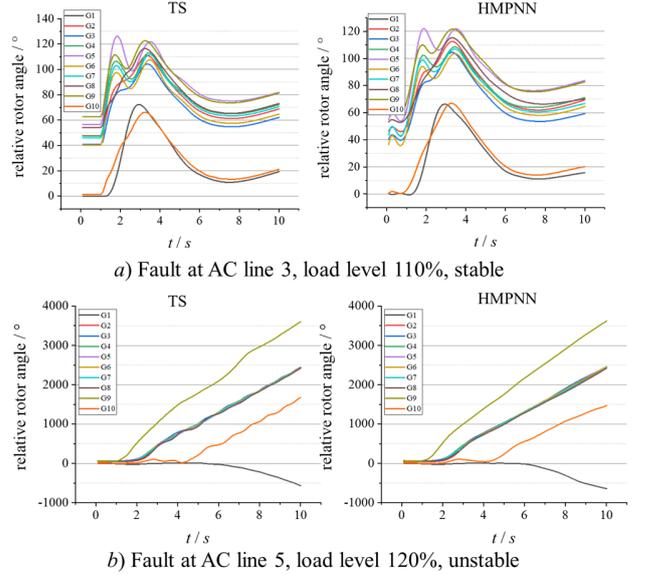

*a)* Fault at AC line 3, load level 110%, stable

*b)* Fault at AC line 5, load level 120%, unstable

Fig. 7 Examples of the HMPNN-based simulator prediction.

## C. Training of the GD2RL

To simulate the uncertainty of renewable sources, the output of PV1 and PV2 are assumed to follow the normal distribution with standard deviations of 30MW and 15 MW, respectively. Although the normal distribution commonly used for PV generation is utilized here, the proposed method does not have constraints on the distribution form and can be applied to any distributions. During the exploration step of GD2RL, a base deterministic state and anticipated contingency is randomly sampled as in VI.A, which also determines the mean output of PV1 and PV2. Then, the outputs of PV1 and PV2 are sampled via MC sampling from the pre-defined distribution. This sampling can lead to power imbalance in the system, which is then distributed among the adjustable generators based on their initial output ratios.

With above sampling, a series of distributional samples along with their corresponding deterministic samples can be obtained, and the number of deterministic samples for each distributional sample is set to 200. For the task networks of the actor and the critic, three-layer neural networks are adopted and the hidden sizes are chosen to be (500, 500, *output_dim*).

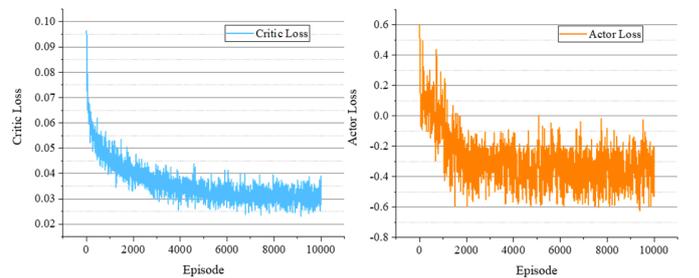

Fig. 8 Loss curves of the GD2RL agent in the training process.

During training, the model is saved every 100 epochs, and the agent that performs the best on the validation set is selected for further case study. The loss curves of the critic network and the actor network in the GD2RL agent during the training process are given in Fig. 8. It can be observed that the loss for the critic and actor during the first few episodes are very high. As the training process goes on, the agent gets lower losses. The negative loss for actor indicates the agent can provide actions to lead the system into a more secure state.

*D. Validation of the redispatch effectiveness of GD2RL*

To test the redispatch effectiveness of the proposed GD2RL agent, distributional samples on the test system are randomly generated for fault on each AC lines. For each fault, 200 distributional samples are generated with high basic power flow level ranges within 100% to 120% (there are few unstable samples in scenarios with low power flow levels). For each distributional sample, 200 deterministic samples are sampled to formulate the distribution. The average confidence is therefore defined as the proportion of stable deterministic samples after redispatch in each fault scenario, and the redispatch confidence of the GD2RL validated by TS is recorded in Table III.

Based on the results in Table III, it can be inferred that the proposed redispatch algorithm exhibits favorable efficiency across diverse uncertain scenarios under different fault conditions, with an average redispatch confidence of 95.41%. Specifically, in scenarios involving uncertain power generation, the confidence surpasses 95% for 17 different fault types. Only a limited number of severe faults, namely five in total, exhibit confidence below 90%, nevertheless, these scenarios still maintain a confidence level above 85%. Considering that the scenarios used for testing are all at high power flow levels, which have relatively low probability of occurring in real-world power systems, the demonstrated performance is quite remarkable.

TABLE III REDISPATCH CONFIDENCE OF GD2RL IN VARIOUS SCENARIOS.

| Fault | Avg. Conf. | Fault | Avg. Conf. | Fault | Avg. Conf. |
|---|---|---|---|---|---|
| 1 | 100.00% | 12 | 97.28% | 24 | 90.56% |
| 2 | 100.00% | 13 | 99.97% | 25 | 92.23% |
| 3 | 93.78% | 14 | 100.00% | 26 | 94.33% |
| 4 | 86.96% | 15 | 100.00% | 27 | 92.05% |
| 5 | 99.67% | 16 | 98.88% | 28 | 94.50% |
| 6 | 99.18% | 17 | 98.27% | 29 | 96.83% |
| 7 | 98.34% | 18 | 94.02% | 30 | 93.24% |
| 8 | 98.78% | 19 | 97.57% | 31 | 93.28% |
| 9 | 96.45% | 20 | 94.10% | 32 | 93.02% |
| 10 | 99.90% | 21 | 87.21% | 33 | 89.32% |
| 11 | 99.90% | 23 | 89.98% | 34 | 88.77% |

Three different distributional samples are used for to visualize the redispatch process of the GD2RL in Fig. 9. To show the impact of the redispatch process on the transient stability of the system, a deterministic scenario is sampled form each distributional sample for the post-contingency rotor angle curve visualization. For each step, the rotor angle curves are obtained by TS and the corresponding TSI is calculated by (26). Case one and two in the figure represents successful cases, while case three represents a scenario that did not achieve successful redispatch within the specified steps (which means that the TSI distribution of this case still does not meet expectations after redispatch). Since our work focuses on the redispatch problem for future scenarios, when the power flow level of the target scenario is high, a successful redispatch cannot be guaranteed solely depending on the adjustment of generator power. Therefore, for those challenging scenarios, the proposed method may not complete the redispatch within the specified adjustment steps, and will signal a redispatch failure to alert the operator to intervene and utilize alternative methods to maintain system stability. From the figure, it can be seen that even for the case where the GD2RL fails, its TSI can be improved, demonstrating that the proposed redispatch method can effectively improve the transient stability of the scenario.

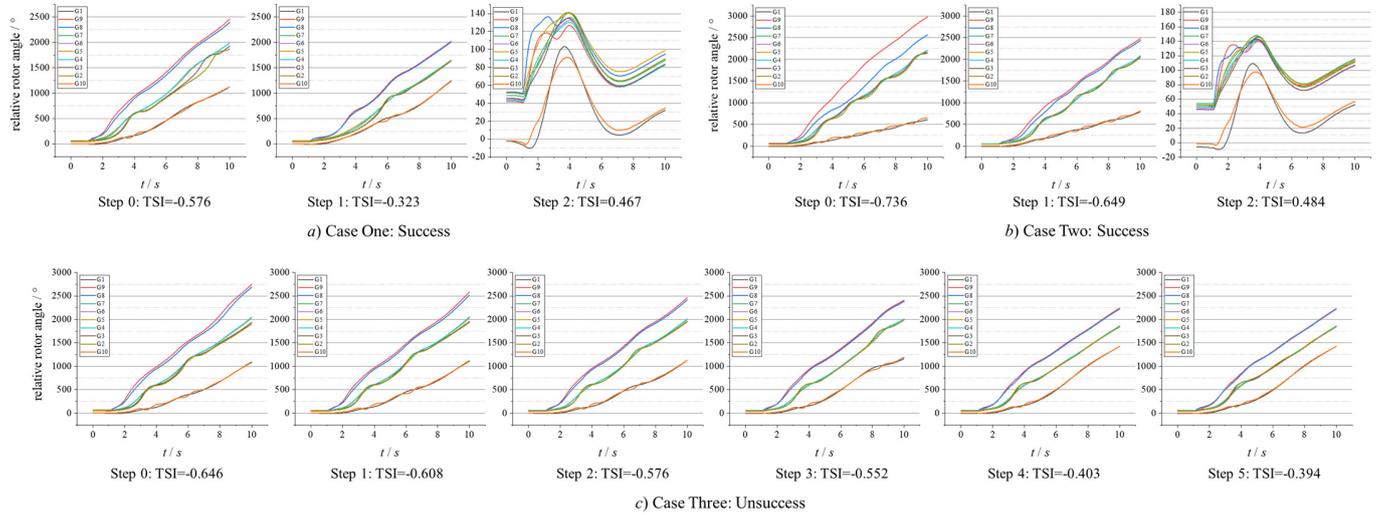

Fig. 9 Visualization of the redispatch process on three different cases.

To further demonstrate that the transient stability of the system is gradually improved when using the proposed redispatch method, the TSI before and after redispatch (marked as $TSI_0$ and $TSI_1$, respectively) in different fault scenarios are calculated via TS. The results are shown in Table IV. Unlike Table III, when calculating the TSI, the deterministic samples that were already stable before redispatch are excluded.

TABLE IV AVERAGE TSI BEFORE AND AFTER REDISPATCH.

| Fault | $TSI_0$ | $TSI_1$ | Fault | $TSI_0$ | $TSI_1$ | Fault | $TSI_0$ | $TSI_1$ |
|---|---|---|---|---|---|---|---|---|
| 1 | -0.51 | 0.51 | 12 | -0.66 | 0.53 | 24 | -0.46 | 0.44 |
| 2 | -0.48 | 0.43 | 13 | -0.72 | 0.56 | 25 | -0.63 | 0.30 |
| 3 | -0.60 | 0.53 | 14 | -0.58 | 0.57 | 26 | -0.66 | 0.31 |
| 4 | -0.52 | 0.30 | 15 | -0.64 | 0.55 | 27 | -0.57 | 0.32 |
| 5 | -0.71 | 0.48 | 16 | -0.63 | 0.53 | 28 | -0.61 | 0.38 |
| 6 | -0.75 | 0.42 | 17 | -0.57 | 0.51 | 29 | -0.63 | 0.42 |
| 7 | -0.76 | 0.47 | 18 | -0.66 | 0.51 | 30 | -0.65 | 0.40 |
| 8 | -0.72 | 0.52 | 19 | -0.60 | 0.54 | 31 | -0.58 | 0.40 |
| 9 | -0.72 | 0.48 | 20 | -0.55 | 0.48 | 32 | -0.65 | 0.36 |
| 10 | -0.71 | 0.55 | 21 | -0.49 | 0.37 | 33 | -0.56 | 0.30 |
| 11 | -0.72 | 0.49 | 23 | -0.48 | 0.39 | 34 | -0.59 | 0.19 |

The average TSI change before and after redispatch in different scenarios is 1.06. The results from Table IV show that the proposed method can effectively enhance the transient stability of the system across different scenarios.

*E. Comparison between GD2RL and PSO*

To validate the superiority of the redispatch strategy obtained from GD2RL, the particle swarm optimization (PSO) algorithm [31], which is a commonly used heuristic algorithm for non-convex optimization, is adopted as a comparison. For each epoch in the PSO, the TS is used to calculate the fitness value for each particle. Since the TS-based transient simulator are embedded in the PSO algorithm, which then approximates local optima through extensive heuristic exploration, the actions obtained through PSO can be can be regarded as the theoretically near-optimal redispatch strategy. Two distributional samples are used to test the difference between GD2RL and PSO, as shown in Table V. The first case suffers transient unstable problem when contingency on AC line 6 occurs and the second case is unstable when contingency is on AC line 28. In the modified New England 39-bus system, the base value of the power per-unit value is 100MW.

TABLE V COMPARISON BETWEEN GD2RL AND PSO ON TWO CASES

| P/p.u. | Init. | Ours | PSO | P/p.u. | Init. | Ours | PSO |
|---|---|---|---|---|---|---|---|
| G1 | 11.13 | -0.36 | -0.28 | G1 | 10.36 | 0 | -0.01 |
| G2 | 5.84 | / | / | G2 | 5.88 | / | / |
| G3 | 7.06 | +0.01 | 0 | G3 | 7.45 | 0 | 0 |
| G4 | 7.43 | 0 | 0 | G4 | 7.26 | 0 | +0.01 |
| G5 | 5.87 | -0.03 | -0.04 | G5 | 5.74 | 0 | +0.03 |
| G6 | 6.86 | -0.03 | +0.08 | G6 | 7.53 | 0 | -0.09 |
| G7 | 6.57 | -0.80 | -0.71 | G7 | 5.82 | -0.99 | -0.50 |
| G8 | 2.95 | 0 | 0 | G8 | 2.23 | 0 | 0 |
| G9 | 3.19 | 0 | 0 | G9 | 4.57 | 0 | 0 |
| G10 | 2.89 | +1.21 | +0.95 | G10 | 2.85 | +0.99 | +0.56 |
| PV1 | 3.18 | / | / | PV1 | 3.89 | / | / |
| PV2 | 4.96 | / | / | PV2 | 5.22 | / | / |
| Time | / | 14.6 s | 9891 s | Time | / | 13.2 s | 9836 s |
| Cost | / | 588 $ | 513 $ | Cost | / | 297 $ | 258 $ |
| Conf. | / | 100% | 100% | Conf. | / | 100% | 100% |

Above results suggest that the redispatch strategies from GD2RL are similar to that from PSO, and the main adjustment directions of the generators are consistent. In both cases, the adjustment cost from GD2RL is slightly higher than that from PSO, but GD2RL significantly reduces the computation time, and both methods achieve 100% confidence level.

To further compare the performance of GD2RL and PSO, 100 testing distributional samples are randomly selected. According to Table VI, the strategies from GD2RL can reach near-optimal average redispatch confidence and cost, while significantly reduce the computational time. If the computation time of PSO is limited within 30 minutes, which is a common online rolling calculation period in practical operating, the proposed GD2RL would yield significantly lower redispatch cost and higher redispatch confidence compared to PSO. By leveraging the rapidity of the GD2RL solution, substantial time savings can be achieved in the problem-solving process. This is of great practical significance, as it allows operators to make timely decisions, ultimately enhancing the overall operational efficiency.

TABLE VI TESTING PERFORMANCE OF GD2RL AND PSO ON 100 CASES

| Index | GD2RL | PSO | PSO (limited) |
|---|---|---|---|
| Mean redispatch cost | 523 $ | 457 $ | 678 $ |
| Avg. Conf. | 95.68% | 95.72% | 90.85% |
| Mean computation time | 13.34 s | 9829 s | 1800 s |

*F. Visualization of the redispatch effect*

To validate the effectiveness of the TSI reward distribution output by GD2RL, three cases with different severity levels are selected for visualization in Fig. 10. For each distributional sample, the final redispatch strategy is the sum of actions for each step. In Fig. 10, the first column shows the pre-redispatch TSI distributions from TS; the second column shows the post-redispatch TSI distributions that are predicted directly by the critic in GD2RL and the third column shows the post-redispatch TSI distributions from TS. The severity of distributional samples A to C increases gradually, which is mainly reflected by their different redispatch difficulties.

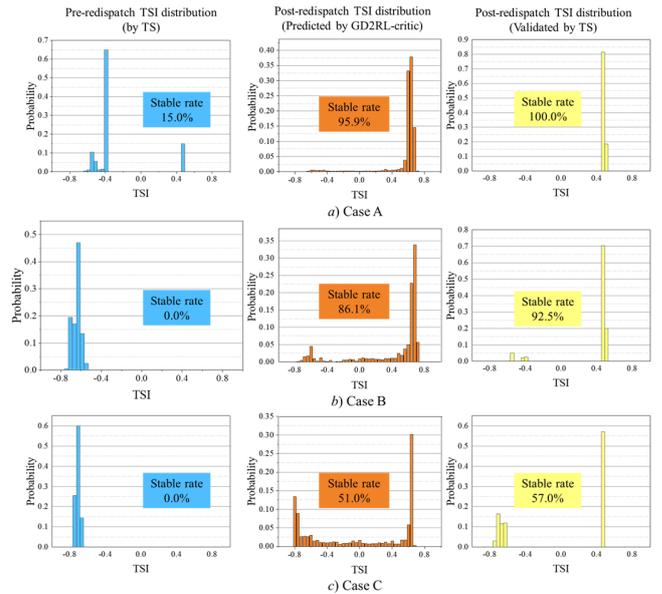

Fig. 10 TSI distributions based on TS and GD2RL-critic.

For case A, 15% of the deterministic samples from the distributional sample are stable. After applying the redispatch strategy generated from trained GD2RL-actor, all deterministic samples are stable with the validation of TS. It can be seen that the GD2RL-critic gives a confidence level of 95.9% for this redispatch strategy, and the TSI reward distribution is quite similar with the distribution from TS. For case B and C, all of the pre-redispatch deterministic samples are unstable, and there is still a certain risk of instability after applying redispatch strategies, at which time the GD2RL-critic can provide the corresponding post-redispatch instability risk, and the distributions of TSI after redispatch are also close to the actual distribution from TS.

Since the proposed GD2RL enables the direct incorporation of prior knowledge regarding the uncertain renewable sources into the RL framework. To validate its advantage, a comparison between DistRL and the traditional RL is conducted, and the details can be seen in Appendix B.

*G. Comparison of the training speed*

In this section, the training time for three different training strategies is compared in Table VII. The first training strategy is using TS as the environment in the DistRL (marked as TS), the second strategy is using HMPNN-based transient simulator to approximate TS in the environment in the DistRL (marked as HMPNN) and the last strategy is including the parallel technique in the exploration step for DistRL (marked as HMPNN+Parallel).

TABLE VII TRAINING TIME FOR DIFFERENT METHODS

| Epoch | TS | HMPNN | HMPNN+Parallel |
|---|---|---|---|
| 1k | 49h 11m | 7h 56m | 2h 13m |
| 2k | 98h 32m | 15h 43m | 4h 20m |
| 3k | / | 23h 40m | 6h 33m |
| 4k | / | 31h 34m | 8h 48m |
| 5k | / | 39h 28m | 10h 59m |
| 6k | / | 47h 25m | 13h 06m |
| 7k | / | 55h 21m | 15h 11m |
| 8k | / | 63h 19m | 17h 21m |
| 9k | / | 71h 11m | 19h 28m |
| 10k | / | 79h 07m | 21h 35m |

It can be observed that the TS is really time-consuming, which takes almost 100 hours for 2k epochs. Due to the long training time for TS-based training, only 2k epochs are recorded. By introducing the HMPNN-based simulator in the DistRL, the training speed can be improved by about five times. On top of that, introducing the distributed parallel technique can further enhance the training speed by three times, which can achieve 10k training epochs within 22 hours.

## VII. CONCLUSIONS & DISCUSSION

In this paper, a novel approach based on graph neural network guided distributional deep reinforcement learning is introduced to address the U-TSCPR problem. By training an HMPNN-based transient simulator, the post-contingency rotor angle curves for generators can be generated efficiently given steady-state and contingency. The simulator is then seamlessly integrated into a D2RL framework, serving as a powerful feature extractor for power system operating states during the environment interaction steps. The proposed GD2RL enables the generation of a deterministic preventive redispatch strategy that carefully considers both security and economy in uncertain operating scenarios. The effectiveness of the proposed method is validated through a comprehensive case study on the modified New England 39-bus system. Main conclusions are as follows:

(1) By introducing the DistRL framework, the operational scenarios with uncertainty can be processed, and the proposed approach is capable of efficiently generating near-optimal redispatch strategies and providing post-redispatch TSI distribution information for uncertain scenarios.

(2) The high-quality HMPNN-based transient stability simulator effectively approximates the TS-based simulator during the RL exploration stage, which yields remarkable enhancements in the training speed. Moreover, by introducing the distributed parallel technique, an additional training acceleration is achieved.

The proposed GD2RL provides a new idea to solve the U-TSCPR problem. Although GD2RL has advantages, there are still some topics worth to be discussed. Firstly, mechanism-based adjustable generator selection can be incorporated into the algorithm to further reduce the dimension of the action space. Secondly, since the graph neural network model is used, it is worth investigating the generalization capability of the proposed method in scenarios with diverse network topologies. Lastly, from a theoretical standpoint, the proposed method can be extended to address redispatch problems that encompass various stabilities, which opens up a valuable avenue for future research.

## APPENDIX A

### THE D4PG-BASED TRAINING ALGORITHM OF GD2RL

**Algorithm 1: D4PG for GD2RL Training**

**Input:** batch size $M$, number of parallel processes $n$, target update rate $\varepsilon$, number of episode $E$, start training threshold $R$
**Input:** number of steps in each episode $T$, discount coefficient $\gamma$

1: Initialize network weights $(\theta, \omega)$ at random, initialize target weights $(\theta', \omega') \leftarrow (\theta, \omega)$, initialize replay buffer
2: Launch $n$ environments
3: **for** $i = 1, \ldots, E$ **do**:
4:    Randomly and parallelly sample $n$ distributional samples $\mathbf{S}_{1\ldots n}$
5:    **if** $i < R$ **do**: Randomly generate $n$ actions $\mathbf{a}_{1\ldots n}$
6:    **else do**: Generate action from the actor: $\mathbf{a}_j = \pi_\theta(\mathbf{S}_j)$ for $j$ in $1, \ldots, n$
7:    **for** $t = 1, \ldots, T$ **do**:
8:      Parallelly execute actions $\mathbf{a}_{1\ldots n}$, observe TSI reward distributions $Z^{TS}_{1\ldots n}$, power flow reward $Z^{PF}_{1\ldots n}$, next state $\mathbf{S}'_{1\ldots n}$
9:      Store transition $(\mathbf{a}_j, \mathbf{S}_j, Z^{TS}_j, Z^{PF}_j, \mathbf{S}'_j)$ in replay buffer for $j$ in $1, \ldots, n$
10:    **end for**
11:    **if** $i \geq R$ **do**:
12:      Sample a random minibatch of $M$ transitions as $\rho$ from replay buffer via a mixed sampling strategy
13:      Update critic by: $\min L(\omega) = \mathbb{E}_\rho \left[ D_{KL}\left( Z^{TS}_j \| \mathcal{T}_{\pi_{\theta'}} Z^{TS}_{\omega'}(\mathbf{S}'_j, \pi_{\theta'}(\mathbf{S}'_j)) \right) \right] + \mathbb{E}_\rho \left[ \left\| Z^{PF}_j - \mathcal{B}_{\pi_{\theta'}} Z^{PF}_{\omega'}(\mathbf{S}'_j, \pi_{\theta'}(\mathbf{S}'_j)) \right\|^2 \right]$
14:      Update actor by: $\min L(\theta) = \mu \mathbb{E}_\rho \left[ \pi_\theta(\mathbf{S}_j) \cdot \mathbf{c}^\top \right] - \mathbb{E}_\rho \left[ \mathbb{E}(Z^{TS}_\omega(\mathbf{S}_j, \pi_\theta(\mathbf{S}_j))) + Z^{PF}_\omega(\mathbf{S}_j, \pi_\theta(\mathbf{S}_j)) \right]$
15:      Update the target networks by soft updating: $\theta' \leftarrow \varepsilon\theta + (1-\varepsilon)\theta'$, $\omega' \leftarrow \varepsilon\omega + (1-\varepsilon)\omega'$
16: **end for**
17: **return** $(\theta, \omega)$

In this work, $M=64$, $n=10$, $\varepsilon=0.001$, $E=10000$, $R=500$, $T=5$, $\gamma=0.99$, and both the critic and actor are updated by a learning rate of 0.001.

## APPENDIX B

## A COMPARISON BETWEEN DISTRL AND TRADITIONAL RL

An analysis between agents based on DistRL and traditional RL is conducted in this appendix. The changes in redispatch confidence as the number of sampled points increased are recorded and the results are shown in Fig. 11.

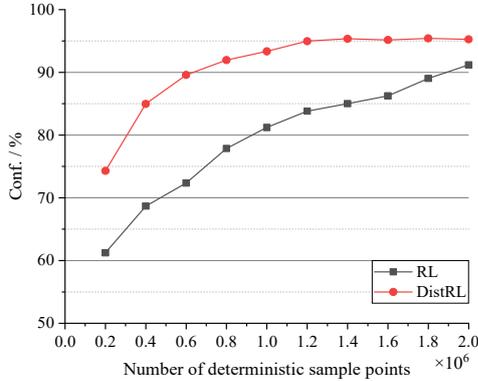

Fig. 11 Redispatch confidence of RL and DistRL with different number of deterministic sample points.

When calculating the reward for each step in the RL, the action generated for the basic scenario will be applied to a deterministic state which is sampled around the basic scenario according to the preset renewable source distribution. For the DistRL agent, one distributional sample contains 200 deterministic sample points.

The results in Fig. 11 clearly demonstrate that, when the number of sampled points is consistent, the DistRL agent achieve notably higher confidence than the RL agent. Moreover, the redispatch confidence curve of the DistRL agent converges more rapidly, while the RL agent fail to converge even after two million sampled points. These findings highlight the fact that the RL agent necessitate an extensive amount of sampling to acquire relatively optimal redispatch strategies, and the proposed DistRL agent has higher learning efficiency.

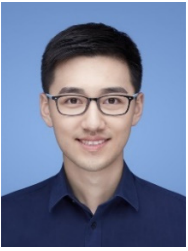

**Zhengcheng Wang** received the B.S. degree from the Department of Electrical Engineering, Tsinghua University, Beijing, China, in 2020. He is currently pursuing the Ph.D. degree in the Department of Electrical Engineering, Tsinghua University. His current research interests include the artificial intelligence and its application in the operation of the complex power system.

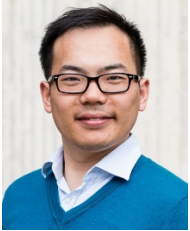

**Fei Teng** (Senior Member, IEEE) received the B.Eng. degree in electrical engineering from Bei-hang University, China, in 2009, and the M.Sc. and Ph.D. degrees in electrical engineering from Imperial College London, U.K., in 2010 and 2015, respectively, where he is currently a Senior Lecturer with the Department of Electrical and Electronic Engineering. His research focuses on the power system operation with high penetration of Inverter-Based Resources (IBRs) and the Cyber-resilient and Privacy-preserving cyber-physical power grid.

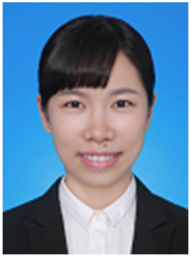

**Yanzhen Zhou** received the B.S. degree from the Dalian University of Technology, China, in 2012, and her Ph.D. degree from Beijing Jiaotong University, China, in 2017. She was a postdoctoral researcher in Tsinghua University from 2017 to 2020. And now she works in department of electrical engineering in Tsinghua University. Her research interests include power system stability and control, and the application of the artificial intelligence technology in power systems.

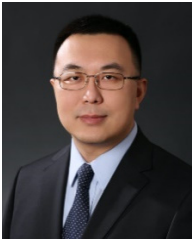

**Qinglai Guo** (SM'14) was graduated from the Department of Electrical Engineering, Tsinghua University, Beijing, China, in 2000 with B.S degree. He received his PhD degree from Tsinghua University in 2005 where he is now a professor. He is now a Fellow of IEEE and IET, and a CIGRE member. Now he is TCPC of Energy Internet Coordinating Committee of IEEE PES, the co-chair of IEEE PES Work Group on "Energy Internet", IEEE PES Task Force on "Cyber-Physical Interdependence for Power System Operation and Control", and IEEE PES Task Force on "Voltage Control for Smart Grid". His research interests include smart grids, cyber-physical systems and electrical power control center applications.

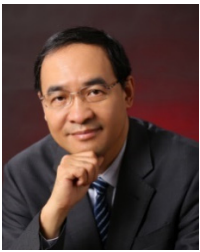

**Hongbin Sun** (SM'12-F'18) received his double B.S. degrees from Tsinghua University in 1992, the Ph.D from Dept. of E.E., Tsinghua University in 1997. He is currently a Changjiang Scholar Chair Professor and the director of Energy Management and Control Research Center in Tsinghua University. He is now a Fellow of IEEE and IET. He is serving as a chair of IEEE smart grid voltage control task force and IEEE Energy Internet working group. He served as the founding Chair of IEEE Conference on Energy Internet and Energy System Integration in Nov 2017. His research interests include energy management system, automatic voltage control, Energy Internet, energy system integration and the application of the artificial intelligence technology in power systems.